\documentstyle[epsf,12pt]{article}
\def\photonatomright{\begin{picture}(3,1.5)(0,0)
                                \put(0,-0.75){\tencircw \symbol{2}}
                                \put(1.5,-0.75){\tencircw \symbol{1}}
                                \put(1.5,0.75){\tencircw \symbol{3}}
                                \put(3,0.75){\tencircw \symbol{0}}
                      \end{picture}
                     }

\def\photonatomup{\begin{picture}(1.5,3)(0,0)
                             \put(-0.75,3){\tencircw \symbol{3}}
                             \put(-0.75,1.5){\tencircw \symbol{2}}
                             \put(0.75,1.5){\tencircw \symbol{0}}
                             \put(0.75,0){\tencircw \symbol{1}}
                   \end{picture}
                  }

\def\photonright{\begin{picture}(30,1.5)(0,0)
                     \multiput(0,0)(3,0){10}{\photonatomright}
                  \end{picture}
                 }

\def\photonrighthalf{\begin{picture}(30,1.5)(0,0)
                     \multiput(0,0)(3,0){5}{\photonatomright}
                  \end{picture}
                 }

\def\photonuphalf{\begin{picture}(1.5,15)(0,0)
                      \multiput(0,0)(0,3){5}{\photonatomup}
                   \end{picture}
                  }

\def\fermionup{\begin{picture}(1,30)(0,0)
                     \put(0,0){\vector(0,1){15}}
                     \put(0,15){\line(0,1){15}}
               \end{picture}
              }
\def\fermionuphalf{\begin{picture}(1,15)(0,0)
                         \put(0,0){\vector(0,1){7.5}}
                         \put(0,7.5){\line(0,1){7.5}}
                   \end{picture}
                  }

\def\fermionull{\begin{picture}(30,15)(0,0)
                        \put(0,0){\vector(-2,1){15}}
                        \put(-15,7.5){\line(-2,1){15}}
                  \end{picture}
                 }
\def\fermionullhalf{\begin{picture}(15,7.5)(0,0)
                        \put(0,0){\vector(-2,1){7.5}}
                        \put(-7.5,3.75){\line(-2,1){7.5}}
                  \end{picture}
                 }
\def\fermionurr{\begin{picture}(30,15)(0,0)
                        \put(-30,-15){\vector(2,1){15}}
                        \put(-15,-7.5){\line(2,1){15}}
                  \end{picture}
                 }
\def\fermionurrhalf{\begin{picture}(15,7.5)(0,0)
                        \put(-15,-7.5){\vector(2,1){7.5}}
                        \put(-7.5,-3.75){\line(2,1){7.5}}
                  \end{picture}
                 }
\def\fermiondrr{\begin{picture}(30,15)(0,0)
                        \put(0,0){\vector(2,-1){15}}
                        \put(15,-7.5){\line(2,-1){15}}
                  \end{picture}
                 }

\def\fermiondll{\begin{picture}(30,15)(0,0)
                        \put(30,15){\vector(-2,-1){15}}
                        \put(15,7.5){\line(-2,-1){15}}
                  \end{picture}
                 }

\newenvironment{Feynman}[3]{\begin{center}
                            \setlength{\unitlength}{#3 mm}
                            \begin{picture}(#1)(#2)
                            \thicklines
                           }{\end{picture} \end{center}}

\newcommand{\nn}{\noindent}
\newcommand{\nll}{\nonumber \\}

\newcommand{\bq}{\begin{equation}}
\newcommand{\eq}{\end{equation}}
\newcommand{\ba}{\begin{eqnarray}}
\newcommand{\ea}{\end{eqnarray}}

\newcommand{\nobody}{\rule{0ex}{1ex}}
\newcommand{\nobodyfrac}{\frac{\nobody}{\nobody}}

\begin{document}
\thispagestyle{empty}
\thispagestyle{empty}
\onecolumn
\date{}
\vspace{-1.4cm}
\begin{flushleft}
{DESY 95--057 \\}
{LMU 07/95 \\}
{hep-ph/95?????\\}
March 1995
\end{flushleft}
\vspace{1.5cm}
{\LARGE  {
Higgs production in $e^+ e^- \rightarrow l{\bar l} \, q{\bar q}$ at LEP and NLC
\vspace*{1.0cm}
}}
\\
\nn
{\large
Dima Bardin$^{1,2 }$, $\;$ Arnd Leike$^{3}$ $\; $ and $\;$ Tord~Riemann$^1$}
\\
\vspace*{1.0cm}

\nn
$^1$ 
DESY -- Institut f\"ur Hochenergiephysik, Platanenallee 6, D--15738
Zeuthen, Germany
\\
$^2$ 
Bogoliubov Laboratory for Theoretical Physics, JINR,
ul. Joliot-Curie 6, 
RU--141980 Dubna, Moscow Region,
Russia
\\
$^3$  
Sektion Physik der
Ludwig-Maximilians-Universit\"at,
Theresienstr. 37, D--80333 M\"unchen, Germany

\vfill
\vspace*{1.0cm}

\nn
Abstract
\vspace*{.3cm}
\\
\nn
Predictions for Higgs production with
processes of the type
$e^+ e^- \rightarrow \mu{\bar \mu}\;b\bar b $
at LEP~2 and the NLC
are calculated.
Short analytical formulae describe the
double differential distribution in the
invariant masses of the $\mu \bar \mu$ and $b \bar{b}$ pairs.
The total cross section may be got with two numerical integrations.
The various Higgs-background interferences vanish either identically
or are small.
The background contributions depend strongly on cuts applied on
the invariant masses.
\vspace*{1.0cm}
\vfill
\footnoterule
\nn
{\small
{\tt
{\footnotesize email:\ \
bardindy@cernvm.cern.ch, leike@cernvm.cern.ch, riemann@ifh.de\\
\nobody$^3$ Supported by the German Federal Ministry for Research
and Technology under contract No.~05~GMU93P
and
EU contract CHRX--CT--92--0004.
}
} }
\pagebreak
\section{Introduction}
The $e^+ e^-$ colliders LEP~2 and NLC ({\it Next Linear Collider})
allow to search for a light Higgs boson with the off shell Higgsstrahlung
process of figure~1a and its background,
\ba
\label{bj}
e^+\;e^- \rightarrow 
\mu \bar \mu + b\bar b,
\ea
where $b$ stands for a heavy fermion ($b,c,\tau$)
with sizeable coupling to the Higgs and $\mu$ for a different one
($\mu,u,d,s$).
Among these final states, $\mu \bar \mu + b\bar b$ has a very clean
experimental signature.
In the $\cal SM$ (Standard Model), a light Higgs boson with a mass in the
region
60 GeV $< M_H <$ 120 GeV would decay with a rate of about 85\% $\div$ 65\%
into $b+{\bar b}$.
Much less frequent are other two fermion (2$f$) final states:
$\tau+{\bar{\tau}}$, $c+{\bar{c}}$.
Above $M_H \sim $ 120 GeV, different processes become more important;
these will not be considered here\footnote{
For a review on basics of (mainly on mass shell) Higgs physics we
refer to~\cite{hreview} and references therein.
}.

Besides the Higgs signal, there are additional contributions from other
4 fermion ($4f$) production processes, which are topologically
indistinguishable from Higgs production
and will be called background.
They proceed via diagrams with
an intermediate $ZZ, Z\gamma$, or $\gamma\gamma$ state (figure~1b) or through
the diagrams of figure~\ref{ldeersd1}.
Further background exists if the final state contains an $e^+e^-$ pair
or neutrinos.
If the two light final state fermions are light quarks, which produce
hadronic jets one has to add incoherently also the production
of the $b\bar b$ pair together with two gluonic jets.

In this letter, we present a complete analytical study of the
invariant mass distribution of~(\ref{bj}) with the
following observable final states:
\begin{itemize}
\item[(i)]
[$\mu \bar\mu, \tau \bar\tau$];
\item[(ii)]
[$l\bar l,b \bar b$], $l =\mu,\tau$;
[$\sum q\bar q,\tau {\bar \tau}$];
\item[(iii)]
[$\sum q\bar q,b \bar b$], $q\neq b$;
[$\sum q\bar q,c \bar c$], $q\neq c$.
\end{itemize}
The pure background cross section has been studied in~\cite{BLR} and
few numerical results for the complete process may be found
in~\cite{zhteu}.
The Monte Carlo approach has been used in~\cite{bargerboosmpittau}.

In the next section, we will calculate the off shell Higgs cross
section.
In section~3, some numerical results are discussed.

\vfill

\begin{figure*}[htbp]
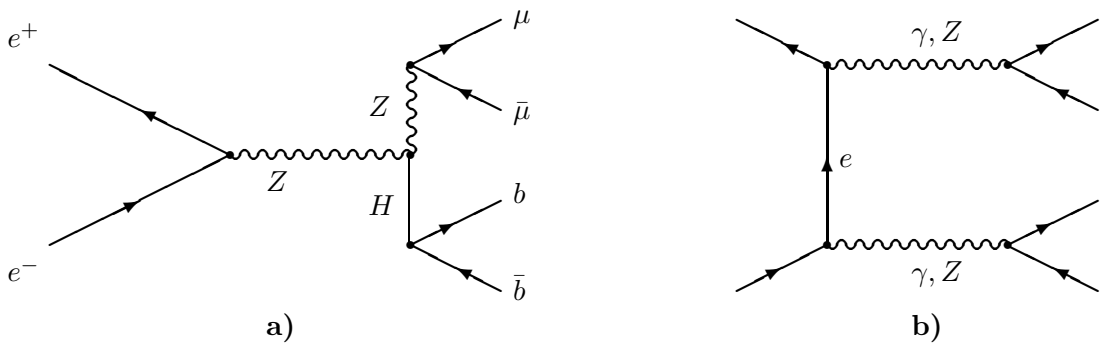

\begin{minipage}[bht]{7.8cm}{
\begin{center}
\begin{Feynman}{75,60}{0,0}{0.8}
%
\put(30,30){\fermionurr}
\put(30,30){\fermionull}
\put(30,30){\photonright}
\put(30,30){\circle*{1.5}}
\put(60,30){\circle*{1.5}}
\put(60,15){\circle*{1.5}}
\put(60,45){\circle*{1.5}}
\put(75,7.5){\fermionullhalf}
\put(75,22.5){\fermionurrhalf}
\thinlines
\put(59.75,15){\line(0,1){15}}
\thicklines
\put(60,30){\photonuphalf}
\put(75,37.5){\fermionullhalf}
\put(75,52.5){\fermionurrhalf}
\small
\put(-07,48){$e^+$}  
\put(-07,09){$e^-$}
\put(36,24){$Z$}
\put(53,36.5){$Z$}  
\put(53,20.0){$H$}
\put(77,06){${\bar b}$}  
\put(77,22){$ b$}
\put(77,36){${\bar \mu}$}  
\put(77,52){$\mu$}
\put(36,00){\bf a)}
\normalsize
\end{Feynman}
\end{center}
}\end{minipage}
\begin{minipage}[bht]{7.8cm} {
\begin{center}
\begin{Feynman}{75,60}{0,0}{0.8}
%
\put(30,15){\fermionurrhalf}
\put(30,45){\fermionullhalf}
\put(30,15){\fermionup}
\put(30,45){\photonright}
\put(30,15){\photonright}
\put(30,45){\circle*{1.5}}
\put(30,15){\circle*{1.5}}
\put(60,15){\circle*{1.5}}
\put(60,45){\circle*{1.5}}
\put(75,7.5){\fermionullhalf}
\put(75,22.5){\fermionurrhalf}
\put(75,37.5){\fermionullhalf}
\put(75,52.5){\fermionurrhalf}
\small
\put(32,28){$e$}
\put(44,49){$\gamma,Z$}  
\put(44,09){$\gamma,Z$}
\put(44,00){\bf b)}
\normalsize
\end{Feynman}
\end{center}
}\end{minipage}
\caption
{
\label{higgs}
\it
The Feynman diagram for off shell $ZH$ production~(a)
and one of the two background diagrams of the {\tt crab} type~(b).
}
\end{figure*}
\newpage

%
\begin{figure}[bhtp]
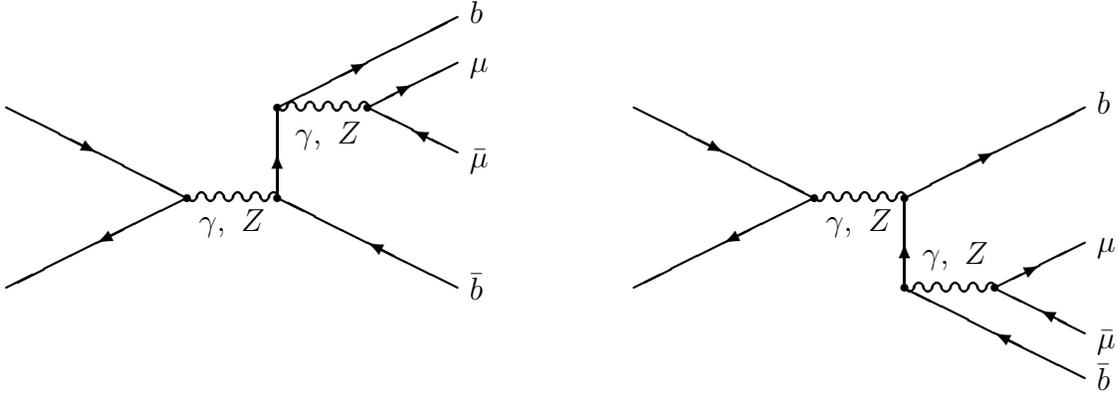

\begin{minipage}[bht]{7.8cm}{
\begin{center}
\begin{Feynman}{75,60}{-5.0,0}{0.8}
%
\put(-10,45){\fermiondrr}
\put(-10,15){\fermiondll}
\put(20,30){\photonrighthalf}
\put(65,15){\fermionull}
\put(20,30){\circle*{1.5}}
\put(35,30){\circle*{1.5}}
\put(35,45){\circle*{1.5}}
\put(50,45){\circle*{1.5}}
\put(35,30){\fermionuphalf}
\put(65,60){\fermionurr}
\put(35,45){\photonrighthalf}
\put(65,52.5){\fermionurrhalf}
\put(65,37.5){\fermionullhalf}
\put(22,24.5){$\gamma,\ Z$}
\put(38,39){$\gamma,\ Z$}
\put(67,59){$ b$}
\put(67,51){$ \mu$}
\put(67,35){${\bar \mu}$}
\put(67,13){${\bar b}$}
\end{Feynman}
\end{center}
}\end{minipage}
\begin{minipage}[bht]{7.8cm}{
\begin{center}
\begin{Feynman}{75,60}{0,0}{0.8}
%
\put(0,45){\fermiondrr}
\put(0,15){\fermiondll}
\put(30,30){\photonrighthalf}
\put(30,30){\circle*{1.5}}
\put(45,30){\circle*{1.5}}
\put(45,15){\circle*{1.5}}
\put(60,15){\circle*{1.5}}
\put(45,15){\fermionuphalf}
\put(75,00){\fermionull}
\put(75,45){\fermionurr}
\put(45,15){\photonrighthalf}
\put(75,22.5){\fermionurrhalf}
\put(75,07.5){\fermionullhalf}
\put(32,24.5){$\gamma,\ Z$}
\put(48,19){$\gamma,\ Z$}
\put(77,44){$ b$}
\put(77,21){$ \mu$}
\put(77, 5){${\bar \mu}$}
\put(77,-2){${\bar b}$}
\end{Feynman}
\end{center}
}\end{minipage}
\vspace*{.2cm}
\caption{\it
\label{ldeersd1}
The $b$-{\tt{deer}} diagrams. The $l$-{\tt{deers}} may be obtained by
interchanging the leptons with the quarks.
}
\end{figure}
\vspace*{0.0cm}
\section{The formulae}
After integrations over four fermion angles and the production angle
of one of the virtual bosons, the following phase space integrals remain
to be performed:
\ba
\sigma(s) =
\int_{
4m_l^2
}^{
(\sqrt{s}-2m_b)^2
}ds_Z
\int_{
4m_b^2
}^{(\sqrt{s}-\sqrt{s_Z})^2} ds_H
{}~\frac{\sqrt{\lambda}}{s}
\frac{\sqrt{\lambda(s_Z,m_{\mu}^2,m_{\mu}^2)}}{s_Z}
\frac{\sqrt{\lambda(s_H,m_b^2,m_b^2)}}{s_H}
\frac{d^2\sigma(s,s_Z,s_H)}{ds_Zds_H},
\label{tsig}
\ea
with the usual definition of the $\lambda$ function,
$\lambda(a,b,c) = a^2+b^2+c^2-2ab-2ac-2bc$, and
$\lambda \equiv \lambda(s,s_Z,s_H)$.
The $s_Z$ and $s_H$ are the virtualities of the $\mu{\bar \mu}$ and $b
{\bar b}$ pairs, correspondingly.
The cross section consists of several pieces:
\ba
\label{tsigd}
\frac{d^2\sigma(s,s_Z,s_H)}{ds_Zds_H}
&=&
\sigma_H +
\sigma_{He} +
\sigma_{H\mu}  +
\sigma_{Hb} +
\sigma_{backgr}.
\ea
Index $H$ means the squared Higgs signal diagram, while index
$Hf$ (with $f=e,\mu,b$) denotes the interference of the signal diagram with
the {\tt crabs} (or conversion diagrams) and the $\mu$- and
$b$-{\tt{deers}} (or annihilation diagrams).
The background cross section $\sigma_{backgr}=
\sigma_e + \sigma_{\mu} + \sigma_{b}  + \sigma_{e\mu} + \sigma_{\mu b} +
\sigma_{be}$
is not related to the Higgs boson and has been studied in~\cite{BLR}.
Again, index $f$ denotes the square of the sum of $f$-{\tt{deers}} and
index $f_1f_2$ the corresponding interferences.
The matrix elements have been squared and integrated over the
five angular variables by two independent calculations with use of
{\tt FORM} and {\tt CompHEP}~\cite{form}.

The off shell Higgs signal $\sigma_H$ reads:
\ba
\label{sig7}
\sigma_H
&=&
\frac{d^2\sigma_H(s;s_Z,s_H)}{ds_Z ds_H}
\nll
&=&
\frac{1}{\pi s} 
\, {\cal T}_H \,
 {\cal C}_{222}(b,s_H;e,s;\mu,s_Z) {\cal G}_{222}(s_H;s,s_Z)
,
\ea
with:
${\cal T}$ -- a color factor, $\cal C$ -- a coefficient function
containing neutral boson propagators and
boson fermion couplings, and  $\cal G$ -- a kinematic function, which depends
only on $s$ and the two virtualities:
\ba
\label{c222}
{\cal T}_H
&=&
N_c(b) \, N_c(\mu),
\\
{\cal C}_{222}(b,s_H;e,s;\mu,s_Z)
&=&
\frac{2}{(6\pi^2)^2} \, m_b^2 \, C_H^4
\frac{1}{|D_{Z}(s)|^2
|D_{Z}(s_Z)|^2
|D_{H}(s_H)|^2}
\nll & &
\times~\left[ L(e,Z)L(e,Z)+R(e,Z)R(e,Z)\right]
\nll & &
\times~\left[ L(\mu,Z)L(\mu,Z)+R(\mu,Z)R(\mu,Z)\right],
\\
\label{g222}
 {\cal G}_{222}(s_H;s,s_Z)
&=&
\frac{1}{4} s_H \left(\lambda + 12 s s_Z\right).
\ea
The conventions for the left- and right-handed couplings
between vector bosons and fermion $f$ are:
$L(f,\gamma)=R(f,\gamma)= (e Q_f)/2$,
$L(f,Z)=e/(4 s_W c_W)\times (2 I_3^f - 2 Q_fs^2_W )$,
$R(f,Z) = e/(4 s_W c_W)$ $\times (-2 Q_f s^2_W)$.
Further, $C_H=e/(4s_Wc_W)$.
We  use in weak amplitudes $e=\sqrt{4\pi\alpha(2M_W)}$,
$Q_e=-1$, $I_3^e=-\frac{1}{2}$, and
$\alpha(2M_W)$=1/128.07  
and define
$s_W^2$=$\pi$$\alpha(2M_W)$/($\sqrt{2}$$M_W^2$ $G_F$).
For photon propagators, $e(\sqrt{s_Z})$ is used instead.
Color factors are $N_c(f)$=1 (3) for leptons (quarks).
The numerical input for the figures is $G_{F} = 1.16639 \times
10^{-5}$ GeV$^2$, $M_Z = 91.1888$ GeV,
$M_W = 80.230$ GeV,
$\Gamma_Z=2.4974$ GeV, $m_b=4.8$ GeV,
and $D_B(s)=(s-M_B^2+i\sqrt{s} \, \Gamma_B(s))^{-1}$.

Another notion exhibits the on shell limit more explicitly~\cite{zhteu}:
\ba
\sigma_H
&=&
\frac{\sqrt{2}G_{\mu}M_Z^2}{s} \frac{M_Z^2}{ss_Zs_H}  {\cal
  G}_{222}(s_H;s,s_Z)
\rho_{Z\rightarrow e^+e^-}(s) \,
\rho_{Z\rightarrow \mu{\bar \mu}}(s_Z) \,
\rho_{H\rightarrow b{\bar b}}(s_H),
\label{zzh}
\ea
where
$
\rho_{B\rightarrow f{\bar f}}(s) = \sqrt{s}\,  \Gamma_{B\rightarrow f
{\bar f}}(s) /(\pi |s-M_B^2+iM_B\Gamma_B(s)|^2)$.
For the Higgs width, we use the Born formula
$\Gamma_H(s)$ = $\sqrt{s}$ $G_F/(4\pi\sqrt{2})$ $\sum_f$ $N_c(f)m_f^2$
in our numerics. In the Higgs mass range considered here this is a
sufficiently good approximation.
The on shell limit is:
$\lim_{\Gamma_B \rightarrow 0} \rho_B(s) = \delta(s -
M_B^2)$BR($B\rightarrow f{\bar f}$).

The $H$--$b$-{\tt{deer}} interference is:
\ba
\sigma_{Hb}
&=&
\frac{d^2\sigma_{Hb}(s;s_Z,s_H)}{ds_Zds_H}
\nll
&=&
\frac{1}{\pi s}
{\cal T}_{H}
\left[\nobodyfrac
{\cal C}_{322}(b,s_H;e,s;\mu,s_Z) {\cal G}_{322}(s_H;s,s_Z)
+
{\cal C}^a_{322}(b,s_H;e,s;\mu,s_Z) {\cal G}^a_{322}(s_H;s,s_Z)\right]
,
\nll
\label{higgsint}
\ea
with the  coupling functions
\ba
\label{c223}
{\cal C}_{322}(b,s_H;e,s;\mu,s_Z)
&=&
\frac{2}{(6\pi^2)^2} \,  m_b^2 \, C_H^2
\nll
&&\times~\Re e \sum_{V_1,V_2=\gamma,Z}
\frac{1}
{
D_{Z}(s)   D_{Z}(s_Z)  D_{H}(s_H)
D_{V_1}^*(s)        D_{V_2}^*(s_Z)
}
\nll & &
\times~
\left[ L(e,Z)L(e,V_1)+R(e,Z)R(e,V_1)\right]
\nll & &
\times~
\left[ L(\mu,Z)L(\mu,V_2)+R(\mu,Z)R(\mu,V_2)\right]
\nll & &
\times~
\left[ L(b,V_1)L(b,V_2)+R(b,V_1)R(b,V_2)\right],
\\
{\cal C}^a_{322}(b,s_H;e,s;\mu,s_Z)
&=&
\frac{2}{(6\pi^2)^2} \,  m_b^2 \, C_H^2
\nll &&
\times~ \Re e\sum_{V_1,V_2=\gamma,Z}
\frac{1}
{
D_{Z}(s) D_{Z}(s_Z) D_{H}(s_H)
D_{V_1}^*(s) D_{V_2}^*(s_Z)}
\nll & &
\times~
\left[ L(e,Z)L(e,V_1)+R(e,Z)R(e,V_1)\right]
\nll & &
\times~
\left[ L(\mu,Z)L(\mu,V_2)+R(\mu,Z)R(\mu,V_2)\right]
\nll & &
\times~
\left[ L(b,V_1)-R(b,V_1)\right]\left[ L(b,V_2)-R(b,V_2)\right],
\ea
and the kinematical functions
\ba
 {\cal G}_{322}(s_H;s,s_Z) &=&
-2ss_Z
\left[\nobodyfrac
(s+s_Z-2s_H)
{\cal L}(s_H;s,s_Z)
+ 2 \right],
\\
 {\cal G}^a_{322}(s_H;s,s_Z) &=&
- s_H
\left[\nobodyfrac
4ss_Z
{\cal L}(s_H;s,s_Z)
+ s_H-s-s_Z \right] .
\label{mutah}
\ea
Further, a logarithm arises from the integration over the fermion propagator:
\bq
{\cal L}(s_H;s_Z,s) =
\frac{1}{\sqrt{\lambda}} \, \ln \frac{s_H-s_Z-s+\sqrt{\lambda}}
                                     {s_H-s_Z-s-\sqrt{\lambda}}.
\eq
All $\cal C$- and $\cal G$- functions are symmetrical in their
last two arguments.
The functions ${\cal C}_{233}$  and ${\cal C}^a_{223}$
get their dependences on $m_b^2$ both from the Higgs coupling
to the $b$ quarks and from the $b$ quark trace.

The $b$-{\tt{deers}} with light quarks have gluon exchange contributions.
Their interference with the Higgs signal vanishes due to the color
trace.

\bigskip
The remaining two types of interferences vanish,
\ba
\sigma_{He}
=
\sigma_{H\mu}
=
0.
\label{intem}
\ea
In these two cases the
$b$ quark trace is odd in the $b$ quark momenta:
$
\mbox{Tr} [ ({\hat p}_b+m_b)$$({\hat p}_{\bar b}$$-$$m_b)$$\gamma^{\alpha}
$$(v_b+a_b\gamma_5)]
$$=$$
4m_bv_b(p_{\bar b}-p_{b})^{\alpha},
$ 
while the rest of the matrix element squared is independent of the $b$ quark
momenta.
Since the
angular phase space integration is symmetric in $b$ and $\bar b$
as may be conveniently seen in the rest system of the $b\bar b$ system,
the result of the integration is zero.
\section{Results and discussion}
%
%
The Higgs mass $M_H$ is unknown.
In the numerical examples it will be varied between 80 and 120 GeV.
Within these bounds, the Higgs is extremely narrow; in the $\cal SM$ one
expects $\Gamma_H < 10$ MeV.
Figure~\ref{f2} contains the total cross sections $\sigma_T$
for the production of
$\mu {\bar \mu} b {\bar b}$,
$(\sum q{\bar q}) b {\bar b}$,
$(\sum q{\bar q})\tau {\bar \tau}$
as functions of the centre of mass energy $\sqrt{s}$ from LEP~2 until
NLC energies.
The Higgs mass is assumed to be $M_H=80$ and, for one of the curves,
120 GeV.

\begin{figure}[bhtp]
\begin{center}
  \vspace{-0.5cm}
  \hspace{-2.3cm}
  \mbox{
  \epsfysize=9cm
  \epsffile[0 0 500 500]{hsqs.ps}
  }
\end{center}
 \vspace{-1.2cm}
\caption[]{\it
\label{f2}
Total cross section $\sigma_T(e^+e^-\rightarrow f_1{\bar f}_1f_2{\bar f}_2)$
in $fb$
for several production channels as a function of $\sqrt{s}$.
All (but one) curves are for $M_H=80$ GeV and
$\sqrt{s_Z},\sqrt{s_H}>60$ GeV.
For the lower lying curve for $\mu {\bar \mu} b {\bar b}$ production
these values are set $M_H=120$ GeV and $\sqrt{s_Z}>60,
\sqrt{s_H}>100$ GeV;
$q{\bar q}=u{\bar u}+d{\bar d}+s{\bar s}+c{\bar c}$.
}
\end{figure}

\begin{figure}[thbp]
\begin{center}
  \vspace{-0.5cm}
  \hspace{-2.3cm}
  \mbox{
  \epsfysize=9cm
  \epsffile[0 0 500 500]{hdsh.ps}
  }
\end{center}
 \vspace{-1.2cm}
\caption[]{\it
\label{f1}
The invariant mass distribution ${\bar \sigma}$ = 
$d\sigma / d\sqrt{s_H}$ in $fb$/GeV
for $\mu{\bar \mu} b{\bar b}$ production
as a function of the invariant Higgs mass $\sqrt{s_H}$ at
$\sqrt{s}=190$ GeV for $\sqrt{s_Z}>60$ GeV and $M_H$=80 GeV.
The plus and minus signs indicate the sign of the interference between
the Higgs signal and the background and the arrow at the resonance
curve the peaking values of the module of the interference.
}
\end{figure}

It is nicely seen that the gold-plated channel
($\mu {\bar \mu}b{\bar b}$ production) is not that with the highest
event rate.
At the $HZ$ production threshold (depending on the chosen Higgs masses
at $\sqrt{s}$ = 171 and 211 GeV)
the cross sections rise steeply and fall then asymptotically like
$1/s$.
Around the threshold this behaviour would be spoiled by the photon
exchange diagrams  if
there were not a cut $\sqrt{s_Z}> 60$ GeV.
A stronger cut becomes possible only after the present Higgs mass limit of
about 60 GeV from LEP~1 will be improved.
Pure background (solid curve) has a similar though less pronounced
behaviour with its threshold at 182
GeV.
For $\mu {\bar \mu} b {\bar b}$ production with $M_H = 120$ GeV the
{\cal SM} background is cut away by a dedicated cut on the background of
$\sqrt{s_Z} > 100$ GeV.
This cut will be allowed after a Higgs mass limit of 100 GeV will be
established.

Figure~\ref{f1} shows for $\mu {\bar \mu} b {\bar b}$ production
the invariant mass distribution $d \sigma
/ d\sqrt{s_H}$ as a function of $\sqrt{s_H}$ at LEP~2 for $M_H=80$ GeV.
The extremely sharp Higgs peak
($ M_H / \Gamma_H \sim 10^4$ !)
is more than four orders of magnitude above the
background.
The Higgs-background interference is extremely small, of the order of
$\Gamma_H/M_H$ compared to the signal and has a zero at the Higgs peak
position.
This justifies the general praxis to assume the Higgs
production being on mass
shell although only the decay products may be observed.
The total cross section arises nearly exclusively from the Higgs
peak region.
Lower cuts may substantially reduce the
background.

The cut on $\sqrt{s_Z}$ is of special theoretical interest.
Figure~\ref{f3}
contains, for the $\mu {\bar \mu} b {\bar b}$
channel, the invariant mass distribution $d \sigma
/ d\sqrt{s_Z}$ as a function of $\sqrt{s_Z}$ at LEP~2 for $M_H=80$ GeV.
In fact, only the low energy region is shown.
Non-negligible cross section contributions come from this region
as may be seen from a comparison with figure~\ref{f1} (compare the scales!).
A cut on $s_Z$ helps considerably to limit this noninteresting background.
The origin are the photonic propagators in the off shell cross section, which
cause a behaviour of the background proportional to $1/s_Z$.
This becomes large if the invariant mass of the muon pair is as small
as $\sqrt{s_Z}=2m_{\mu}$.

\begin{figure}[thbp]
\begin{center}
  \vspace{-0.5cm}
  \hspace{-2.3cm}
  \mbox{
  \epsfysize=9cm
  \epsffile[0 0 500 500]{hdsz.ps}
  }
\end{center}
 \vspace{-1.2cm}
\caption[]{\it
\label{f3}
The invariant mass distribution ${\bar \sigma}$ = 
$d\sigma/d\sqrt{s_Z}$ in $fb$/GeV
for $\mu{\bar \mu}b{\bar b}$ production
as a function of the invariant $Z$ mass $\sqrt{s_Z}$ 
for $\sqrt{s}=190 $ GeV and $\sqrt{s_H}>60$ GeV.
}
\end{figure}

Maybe here is the right place that one should mention all the finite mass
effects, which
influence the various cross section parts and which
from time to time become a
point of discussion when different numerical results are compared.
The following mass dependences are taken into account:
\begin{itemize}
\item
The $m_b$ in the Higgs $b$ quark coupling;
\item
The $m_b$ in the trace of the $b$-{\tt{deer}} when interfering with
the signal;
\item
The $m_{\mu}$ in the {\tt deers} if no cut is applied to $s_Z$.
\end{itemize}
The first two items are evident from the formulae of section~2.
The third one deserves a comment.
The product $\sqrt{\lambda(s_Z,m_{\mu}^2,m_{\mu}^2)}$
$(1+2m_{\mu}^2/s_z)$ $D_B(s_Z)$ / $s_Z$ 
contains, besides the propagator, a factor from phase space and one
from the squared matrix elements.
For extremely small $s_Z$, this combination
yields for photon exchange ($B=\gamma$)
a {\it finite correction} to the cross section of order ${\cal O}(1)$
which is erroneously absent if the muon mass is set zero.
(see figure~\ref{f3}, dashed line).
Thus, in no-cut calculations this mass correction is not allowed to be
left out while even a weak cut removes this dependence completely.

Further, we should mention that many other terms
with a dependence on $m_b^2$  or $m_{\mu}^2$
also occur and could be also taken into account; not all of
them are safely smaller than the Higgs-background interference.
Since they are small and not of the leading order, we neglect them all.

A crude estimate of the peak position may be obtained for the on shell cross
section by the following ansatz:
$
\sigma_T^{on}(s)
\sim
C \sqrt{s-(M_Z+M_H)^2} / s^{3/2}
$.
The extremum of this is located at
$
\sqrt{s_{peak}} = \sqrt{3/2} (M_Z+M_H) \sim 1.22 (M_Z+M_H)
$, which is a few dozens GeV away from the threshold and, of course, gets
changed by a refined treatment and the more for the off shell case and
with QED corrections.
{}From figure~\ref{f2} one may see that the peak position depends on the
production channel so that a trivial universal estimation of
$\sqrt{s_{peak}}$ seems not to exist.
Finally,
we also should mention that there are substantial radiative corrections
to $4f$ production~\cite{hreview}.
The discussion of them in~\cite{BLR} applies also here and the Fortran
program {\tt 4fan}, which was mentioned there has been used in order
to produce the figures of this letter.

{\it To summarize,}
we performed the first complete semi-analytical calculation of off mass shell
${\cal SM}$ Higgs production with background
for the process $e^+e^-\rightarrow\mu{\bar \mu}b{\bar b}$.
The importance of the various dependences on $m_b$ and $m_{\mu}$
has been studied.
The Higgs background interferences either vanish identically after integration
over the fermion angles or are extremely small due to the narrow Higgs width.
The background may be reduced substantially by dedicated cuts on both
virtualities $s_H$ and $s_Z$.
For practical purposes this means that Higgs model specialists may calculate
their isolated Higgs signal contributions leaving out the background.
The latter one may be added later incoherently while the Higgs
background interferences may be safely neglected.
These general conclusions apply also to physics beyond the ${\cal SM}$
as long as the Higgs width remains small.

\end{document}